\documentclass[submission,copyright,creativecommons]{eptcs}
\providecommand{\event}{WICSE 2010} 
\usepackage{breakurl}             
\usepackage{graphicx}
\usepackage{xcolor}
\usepackage[T1]{fontenc}
\usepackage[latin9]{inputenc} 

\title{Contract Aware Components, 10 years after}
\author{Antoine Beugnard
\institute{Telecom Bretagne\\ Brest, France}
\email{antoine.beugnard@telecom-bretagne.eu}
\and
Jean-Marc Jézéquel \qquad\qquad Noël Plouzeau
\institute{IRISA\\
Rennes, France}
\email{\quad jezequel@irisa.fr \quad\qquad plouzeau@irisa.fr}
}
\def\titlerunning{Contracts 10 years after}
\def\authorrunning{A. Beugnard, J.M. Jézéquel \& N. Plouzeau}
\begin{document}
\maketitle

\begin{abstract}
  The notion of {\em contract aware components} has been published
  roughly ten years ago and is now becoming mainstream in several
  fields where the usage of software components is seen as
  critical. The goal of this paper is to survey domains such as
  Embedded Systems or Service Oriented Architecture where the notion
  of contract aware components has been influential.  For each of
  these domains we briefly describe what has been done with this idea
  and we discuss the remaining challenges.
\end{abstract}

\section{Introduction}

In a special issue of IEEE Computer published in July~1999, we
published an article entitled {\em Making Components Contract
  Aware}. The goal of this article was to discuss the extension of
Meyer's Design-By-Contract idea to the world of software
components. Our finding were that component contracts would have to
deal with specific concerns that we classified into four categories,
from syntactic to semantic to synchronization, up to Quality of
Service.

There is empirical evidence that this article has got some lasting
impact on the field of Component Based Software Engineering. For
instance, on June 2010 it has been cited 545 times by other papers
(according to Google Scholar). Several of these papers themselves got
quite a sizable number of citations. Impacted domains
include\footnote{Number in parenthesis is the number of citations of
  this paper according to Google Scholar.}:
\begin{description}

\item[Components:] {\em Software reuse strategies and component
    markets} \cite{cacm:2003} (81 citations), {\em Specification,
    implementation, and deployment of components} \cite{cacm:2002} (80
  citations), {\em Interface Compatibility Checking for Software
    Modules} \cite{cav:2002} (78
  citations) 

\item[Adaptation:] {\em Composing Adaptive Software}
  \cite{adapt:computer:2004} (281 citations), {\em A Taxonomy of
    Compositional Adaptation} \cite{adapt:tr:2004} (53
  citations)

\item[Internet (beyond components):] {\em Information agent technology
    for the Internet: A survey} \cite{agent:survey:2001} (199
  citations), {\em WSOL -- Web Service Offerings Language}
  \cite{wsol:2002} (106 citations)

\item[Real-time:] {\em Monitoring, Testing and Debugging of
    Distributed Real-Time Systems} \cite{rt:2000} (90 citations)

\item[High Performance Computing:] {\em A Component Architecture for
    High-Performance Scientific Computing} \cite{hpc:2006} (86
  citations)
\end{description}

This has prompted us to try to reflect upon the evolution of the
contract-aware components domain roughly ten years after the
publication of our article in IEEE Computer (we will refer to this
article as the MCCA paper in the following sections). The goal of the
present paper is thus to highlight domains where the notion of
contract aware components has been influential, to briefly describe
what has been done with this idea and to discuss the remaining
challenges.  The rest of this paper is organized as follows.  In
Section~\ref{originalPaper}, we briefly recall the main ideas
discussed in our original paper. In Section~\ref{related} we give an
overview of the state of the art with respect to contract-aware
components in domains such as Embedded Systems or Service Oriented
Architecture. We conclude the paper with a discussion on the possible
evolutions of the contract concept.

\section{Original paper summary}
\label{sec:summary}\label{originalPaper}

Ten years ago, we proposed in~\cite{computer:1999} to apply B. Meyer's
Design by Contract~\cite{computer:1992} principles to components with
an attempt at generalization. We introduced a classification of
contracts in four levels, and beyond a simple verification use of
contract, the possibility to manage and negotiate contracts at
runtime.

The classification, like all classifications, expresses a point of
view that we discuss a little further (section~\ref{sec:classifications}). It
can be summarized as:

\begin{description}
\item[Syntactic (or basic)] The goal is to make the system work. It is
  generally specified with Interface Definition Languages (IDLs),
  as well as typed object-based or object-oriented languages. It
  ensures the components can be assembled. 
\item[Behavioral] The goal is to specify each operation. It is
  generally specified with a couple of assertions: a precondition and a
  postcondition. It ensures the operations offered and required are
  not only syntactically compatible but also semantically.
\item[Synchronization] The goal is to specify the coordination of
  operations. It can be specified with an automaton labelled with
  operations. It ensures the operations are used in the proper order.
\item[Quality of Service] The goal is to quantify a few features
  associated to operations. Performance, availability and quality of
  result can be specified and negotiated at that level.
\end{description}

This classification clearly helps to structure the specification and
to understand the coverage of requirements a component has.

The structural decomposition that we proposed also introduces a temporal
decomposition with a contract management life-cycle. The main idea is
to reify contracts (defining contracts as objects). The life-cycle can
be summarized as follows:

\begin{description}
\item[Define] Describe the component features with all required
  contracts. A component is considered in isolation.
\item[Subscribe] Select  from all contracts  those that
  are useful in the context of component use, and configure them.
  Components are assembled  (used) with an intent.
\item[Check] Evaluate contracts and react accordingly. The moment the
  checking is carried out depends on the level of contract: usually
  levels 1 and 2 can be statically checked while levels 3 and 4
  require some runtime monitoring. In some cases, level 3 can be
  statically checked. In the case of a contract violation we proposed 4 kinds
  of reactions: ignore, reject, wait or negotiate. 
\item[Terminate] Decide when to stop evaluating contracts. 
\end{description}

The 4-level structural classification is oriented towards the nature
of the constraints that have to be specified. 
As we will see later
(section~\ref{sec:classifications}) space and time information are
other classification dimensions that can be considered.

In the next section, we focus on some application domains in order to
analyze how contracts were used.   

\section{Related work\label{related}}

Contracts have been applied in many fields. They help to design and
structure both the product and the production process. In this
section we focus on examples which range from the less flexible architectures
(closed embedded systems) to the most flexible ones (service oriented
architectures) through an intermediate flexible architecture (component based
systems). This shows that ``contract awareness'' is not limited to
components, but can efficiently be extended to other paradigms such as
services.  

\subsection{Embedded Systems}

In recent years, the notion of contract aware components has clearly
percolated in the domain of Real-Time Embedded Systems. For instance,
the Artist network of excellence supported a research cluster on
Component-Based Software Development where the notion of contract was
central. However, a key characteristic of component-based embedded
systems is the wide heterogeneity of component models. This
heterogeneity can be found in different execution models (synchronous,
asynchronous, vs. timed), different communication models (synchronous
vs. asynchronous), as well as different scheduling
paradigms. Designing heterogeneous embedded systems from diverse types
of components, and allowing the prediction and optimization of
functional and non-functional properties of the designed systems is
still an open challenge~\cite{BasuBS06}.  Thus there is a need to
develop an innovative theory for building complex heterogeneous
systems which better addresses composability and compositionality
issues~\cite{Sifakis09}. Such a comprehensive theory is still missing
today, thereby making it difficult to understand how to build systems
that combine, e.g., synchronously and asynchronously executing
components and reason about non-functional properties, but this is the
subject of very active research (e.g. as published in the Emsoft
conferences).

It is also now acknowledged that a key issue in component-based
embedded software development is about handling non-functional
properties (including real-time and QoS properties). The notion of a
rich component model~\cite{Damm05,BenvenisteCFMPS07} is thus gaining
momentum to help engineers to model, specify, and predict timing, QoS,
and resources properties of components and of systems composed from
components. Still, typical support for handling QoS and resource usage
is rather limited.

Specific non-functional (level 4) contracts are dealt with in ad hoc
manners. For timing properties, different variants of timed automata
have been used, as in, e.g., the Omega component
model~\cite{DammJPV05}. For properties relating to queuing and
performance, models based on queueing networks, Markov chains,
etc. have been used. These approaches offer a precise mechanism for
specifying and analyzing QoS properties, but they suffer from
scalability problems.

In Real-Time Embedded Component-Based Systems, level 3 contracts are
also of prime importance to manage both expectations of the component
about its environment and guarantees offered in return by the
component to its environment. An example of level~3 contracts is the
notion of \emph{interface automata}~\cite{AlfaroHS02}, viewed as
enriched type systems (the so-called \emph{behavioral type systems}),
which capture the ordering aspects of software component
interactions. Interface automata provide a semantically well-founded,
built-in notion of refinement: a component refines another one if it
imposes weaker constraints about the environment and offers stronger
guarantees in return~\cite{cav:2002}.

UML emerged in recent years as a modeling standard for software,
including software for embedded systems for which specific UML
profiles have been developed. A first attempt was made with the UML
Profile for Schedulability, Performance, and Time (SPT) to model
real-time concerns. However, the SPT profile suffered from several
shortcomings, and was quickly superseded by the MARTE profile
(Modeling and Analysis of Real-Time and Embedded systems), which
better addresses issues such as compliance with the UML2.0
specification of not only real-time constraints but also other
embedded QoS characteristics such as memory and power consumption,
modeling and analysis of component-based architectures, and the
capability to model systems in different modeling paradigms
(asynchronous, synchronous, and timed). Still very few tools are
currently able to exploit MARTE to its full extent.

\begin{figure}
  \begin{center}
    \includegraphics{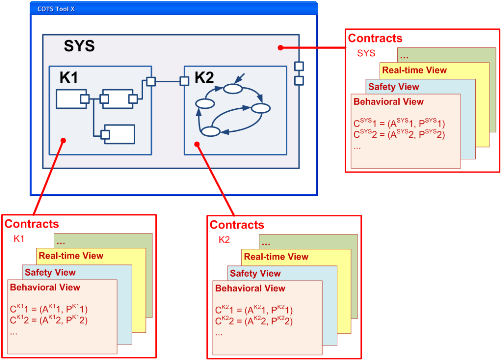}
  \end{center}
  \caption{The SPEEDS vision: Design enriched by contracts}\label{speedsfig}
\end{figure}

In this domain, SPEEDS (SPEculative and Exploratory Design in Systems
Engineering) is another example of a recently completed European
project that developed contracts and multi-viewpoint as core concepts
(see Figure~\ref{speedsfig}).

SPEEDS has delivered a solution based on enriching functional
decomposition with contracts, \emph{i.e.} textual and formal
descriptions of functional and non-functional aspects. The SPEEDS
version of contracts give abstract descriptions of the exposed
properties of a component, classified as \emph{assumptions}
(requirements of the component with respect to its environment) and
\emph{promises} (guaranteed delivered properties of the component
provided its assumptions have been fulfilled). According to the
documentation produced by the SPEEDS
project\footnote{http://www.speeds.eu.com/}, typical contracts
specifying the behaviour of a component (used as promises) could be
for instance:
\begin{itemize}
\item The output \emph{out} is the sum of the inputs \emph{in1} and \emph{in2}.
\item Every request will be served.
\item A request will be served within 10~ms.
\end{itemize}
Conversely, examples of behavioural contracts used as assumptions could be:
\begin{itemize}
\item Input data \emph{in1} will never show a negative value.
\item The data rate of in1 will not exceed one message in 10~ms.
\end{itemize}

SPEEDS has provided a complete framework for modeling, combining,
analyzing and managing such multi-viewpoint, enriched models.

\subsection{Component architectures}

Contracts have gained first class status in recent years.  The set of
concepts used to describe and manage them is now very rich.  More than
ten years ago, the first approaches relied on a static model of
contract. For instance, an operation definition may provide a non
negotiable precondition (the caller cannot negotiate the precondition)
and also a non negotiable postcondition.  While this may be sufficient
for level~2 contracts (especially if one considers level 2 contracts
as a type mechanism) this scheme is too static to maximise reuse and
adaptation to evolving needs.  Hence a negotiation model is
required. In turn this implies that contracts have complex features
and that they are much more than a pair of precondition/postcondition
predicates.

In this section we use the Confract approach~\cite{confract:cclj:2007}
as an example architecture to illustrate elaborate means of contract
management (especially negotiation and monitoring).  Confract is a
contract management framework for the
Fractal~\cite{bruneton2006fractal} component model.  Fractal has a
rich component life cycle, which in turn allows for rich contract
lifecycle with high possibilities regarding dynamic contracts.  As in
most modern contract based systems, a Confract contract is a
negotiated agreement between several parties whose responsibilities
are clearly stated for each contract feature.  Confract allows for
several kind of contracts, depending on the location that receives
them:
\begin{itemize}
\item interface contracts describe agreement between a service
  provider and a service client;
\item external composition contracts express compound usage rules for
  a component, defining how a component can be used globally (such
  contracts encompasses several interfaces);
\item internal composition contracts define how internal components
  are connected to external interfaces.
\end{itemize}

The various contract locations mentioned above show that the
cooperation properties are more central than in the case of initial
\emph{design by contract} approach.

The set of specifications is at the heart of Confract: a contract is a
``collaboration hub''.  Through a set of operations entities
(component instances), the partners select the specification items
that they are interested in (either as service provider of service
consumer) with adequate values for the specification parameters.  Once
partners have reached an agreement on a specification term, this term
is ``closed''.  A closed term includes a contract manager (which is in
charge of managing the violation of this term), contributors (which
strive at ensuring the contract term by providing services) and
beneficiaries (which are customers of the contract bound services).
When all terms are closed, the contract can be instantiated and used.

In a typical architecture, an instantiated contract is a reference to
monitor the activities of the partner and detect contract violations.
This contract instance is at the center of the collaboration
management.  In service based systems and also in evolving component
based systems configurations can be computed on the fly and therefore
contract models must also take these evolutions into account.
Techniques such as model driven engineering help to master
reconfiguration and they play a key role in contract management.  More
precisely, the ``model at runtime'' technique~\cite{Morin09f} allows
for a precise reconfiguration control of a running system.  This
includes the analysis of contract validity on an evolving
architecture.

\subsection{Service Oriented Architecture}

Contract management is central to Service Oriented Architecture
design.  With respect to component based architectures, service
oriented ones are characterized by the highly dynamic configuration
and reconfiguration capabilities that they require.  As in component
based architecture, SOA contracts are the key concept to setup
cooperation between entities.  But SOA contracts have a richer
lifecycle: they are not set in stone at design time but are much more
malleable than design time component contracts.  Conditions that were
favorable to the definition of a cooperation at a given time and place
may change at any moment and this change lead to a deal break.  In
turn this implies a reaction from the contract partners, from a basic
renegotiation of bounds to a contract cancellation.  In the spirit of
service oriented architectures, such contract volatility is not a bug
but an essential feature.

This high level of dynamicity has a profound impact on the way to
create, configure and supervise contract. This leads to a large set of
concepts (see~\cite{keller2003wsla} for an example of such conceptual
framework).  Hereafter we will mention some points on configuration
and monitoring of contracts in a SOA context.

The terminology of the SOA domain includes the notion of service level
agreement (SLA).  There is a connection between the contract notion
(in the sense described in the previous section on components) and the
SLA concept.  Both concepts share the idea of mutual agreement on
provided and consumed services.  However, while component contracts
are based on precise notions of quality, referring to low level items
of execution (e.g. maximum time between service invocation and service
termination events), SLA properties refer to higher level properties,
abstracting many execution details and often using stochastic
descriptions.  SLA deals with the global performance of a system as a
service provider, and SLA descriptions include various metrics
concerns and dimensions, such as business metrics: financial
properties, for instance a billing computation formula based on
stochastic behavior of a compound metrics (e.g. mean execution time of
a service during the last hour, etc).

\paragraph{Contract creation and configuration}

Strictly speaking, contracts are created from offers, once offers have
been read, chosen and configured.  In the component design world,
components are interconnected through ports and in many component
metamodels, the relationship between a provided port and a required
port is bidirectional and represented by a connector (explicitly or
implicitly). A contract is built from a required properties part bound
to the requiring/client port, matched with provided properties bound
to the provider port.

Regarding negotiation aspects, SOA offers a wide range of
possibilities for every contract level, in the sense of initial MCCA
article:
\begin{enumerate}
\item level 1 negotiation is realized by the service discovery
  mechanism inherent to service based systems; the notions of type
  compatibility are managed there;
\item level 2 negotiation is also connected to service discovery: pre
  and postconditions are a form of type definition that extends data
  type definitions;
\item level 3 is realized by the definition of an orchestration;
  usually negotiation is limited because orchestrations are more
  static than the services they rely upon;
\item level 4 is realized by service level agreement mechanisms.
\end{enumerate}

\paragraph{Monitoring services}

This dynamic approach requires that SOA contracts have to be
supervised at all times, contrary to a more static view of contract
composition analysis from the embedded or component systems design.
In practice, the monitoring system is generated and configured by the
contract management infrastructure. Monitoring can be performed by
external parties in order to obtain a neutral point of view of the
compliance of the interested parties to the contract.

With respect to the classification of the MCCA paper, monitors can
manage different contract levels: monitoring systems such as the one
in~\cite{baresi2007towards} rely on an infrastructure that can accept
plugins for various monitoring tactics, with a rule-based engine to
trigger reaction strategies.  Reaction strategies may adapt monitors
and monitoring rules, as well as strategies that retry, rebind or
reorganize.

Regarding monitoring, service based architectures address the four
MCCA levels with very different techniques:
\begin{enumerate}
\item level 1 contracts can be managed by WSDL alike service
  descriptions, with a type system that is loose enough to deal with
  heterogeneity and fast paced evolution; contract compliance is
  computed by loose type computation and checking;
\item level 2 contracts can be described by pre and postcondition and
  can be checked by weaving monitors, inserting adapters, etc~
  \cite{baresi2007towards};
\item level 3 contracts can be addressed by orchestrations, which
  somewhat can ease contract monitoring because the control is
  centralized;
\item level 4 contracts can be addressed by SLA validation and monitoring
\end{enumerate}

Monitoring techniques are different for embedded architectures and
service based systems:
\begin{itemize}
\item service based designs have limited possibilities of static
  analysis tools because of the highly dynamic nature of these
  systems.  Checking can be done at run time only: at contract
  negotiation time, during the execution of service requests, at
  system reconfiguration time;
\item the consequences of contract violation differ between
  traditional component based systems and service based ones. The
  goals of monitoring are therefore different: for embedded
  architectures, monitoring is a complement to more static analysis
  and validation techniques; for service based systems monitoring is a
  first class concept that is an integral part of the contract life
  cycle.
\item monitors for embedded systems can be generated statically and
  injected in the main system code infrastructure to allow for
  efficient violation detections and fast error
  management~\cite{Saudrais07e}; on the contrary dynamic systems such
  as service based architectures are more difficult to generate in
  advance and violation management policies can also be defined
  dynamically.
\end{itemize}

\subsection{Other work}

The initial application field of the MCCA paper was components. A wide
interpretation of the word component is allowed and leads to several
applications of contracts. Moreover, as discussed in
section~\ref{sec:classifications}, a contract can be attached to many
kinds of component: from operation (tiny ``component'') to system
(huge ``component''). In this section we describe briefly other work
using contracts.

\begin{figure}
  \begin{center}
    \includegraphics[scale=0.75]{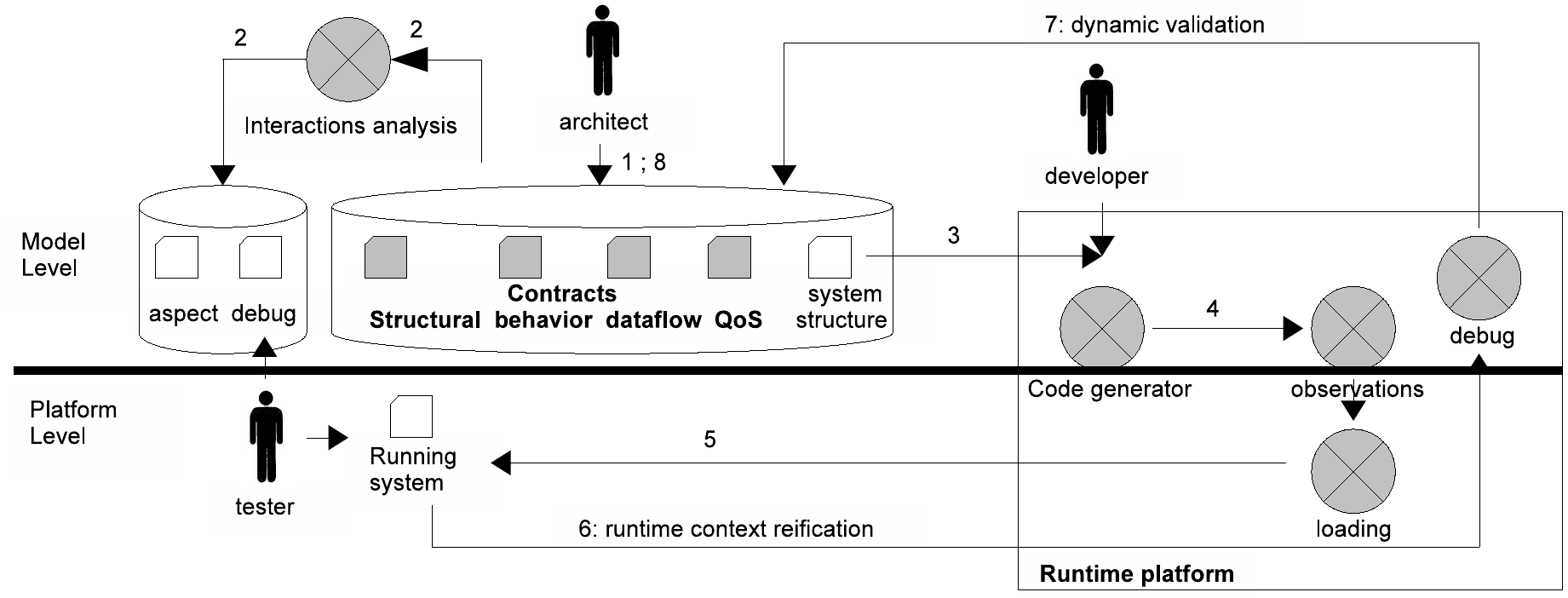}
  \end{center}
  \caption{A full process with contracts from~\cite{waignier}\label{sca}}
\end{figure}

A recent interesting work has been realized by~\cite{waignier}. The author
proposed an agile framework for the evolution of systems based on
components and services. Figure~\ref{sca} shows the whole development
process. Four types of contracts (gray boxes) are used at different stages of the
process with different tools. Some contracts are used with static
checkers in order to prove architectural static properties while other
contracts are transformed and integrated into the code to enable
observations and runtime monitoring. This work shows that contracts can
be a natural interface in order to mix and assembly components and
services as in the Service Component Architecture initiative. 

In 2000, the OMG introduces the MDA (Model Driven Architecture)
initiative~\cite{mda}. Beyond the UML or MOF dependency of this
approach, this initiative promotes a systematic use of models
(whatever the modeling language) and of model transformations
(whatever the transformation language) that describes design
decisions. In~\cite{kabore}, E. Kabor\'e used contracts to specify all
transformations; constraints on input and output models are considered
as pre and postconditions on transformations. 

More recently, A. Koudri proposed Modal~\cite{koudri} a process model
to interpret parts of a software development process as ``process
components'' which, hence, allows to attach contracts to them.

\section{Discussion on the classification}
\label{sec:classifications}

As for any classification, our four levels contract classification
expresses a point of view and can be discussed. The point of view of
the MCCA paper is to consider the component as a whole and not to
consider it as a set of parts such as interfaces, ports or other
``components''. A contract is then attached to a component, but this
position is easily reconsidered with contracts attached to components
parts. Basic and semantic contracts are naturally linked to
operations, themselves grouped into interfaces or ports whatever the
boundary of the component is composed of. Synchronization (or
coordination) and quality of service contracts describe more global
features and are easily attached to the component itself. In his
thesis, G.~Waignier~ \cite{waignier} proposes a systematic approach
using contracts at many structural levels: operation, port, whole
component, assembly of components (this is made possible by the
Confract component model~\cite{confract:cclj:2007})

The levels that are the most subject to interpretation are the
behavior and synchronization ones. This is probably because
synchronization is a part of the component behavior and that languages
used to specify operation semantics in preconditions and
postconditions are rich enough to describe control states. In a strict
interpretation of the behavioral contract, the operation specification
is stateless and makes no assumption on the state of the
component. For instance, a deposit in a BankAccount is \texttt{balance
  += amount;} whatever the BankAccount status (non-existing, open,
closed, deleted, etc.) is. Unfortunately, languages such as OCL allows
references to variables that can encode a state, introducing a
synchronization-level issue in the behavioral level. In order to
remove this moving borderline, the Accord project~\cite{accord}
proposes a classification in only three levels: syntactic, semantic
and pragmatic. But, this usual classification in language theory
introduces a new kind of ambiguity; pragmatic aspects usually denote
the way things are used. Therefore, is the synchronization contract
that describes the way operations are used a pragmatic contract (from
the operation point of view) or a semantic contract (from the
component - as a unit - point of view)?

Finally, a completely different and complementary classification is
based on the moment a contract is designed within the design
process. All contracts are defined and refined during the whole
development process. A more precise classification can define
requirement contracts, specification contracts, design contracts,
implementation contracts and runtime contracts.  UML
Components~\cite{umlcomponents} and G. Waignier~\cite{waignier} rely
on such an approach. 

As a conclusion for this discussion on contract classifications we can
distinguish three dimensions :
\begin{enumerate}
\item the nature of the contract, as in the MCCA initial proposal
\item the location of the contract;
\item the process moment (time position) of the contract.
\end{enumerate}

The nature of a contract depends on its attachment and its position in
the development cycle of a component. All these contracts have
compatibility constraints that need to be analyzed along the
process. The classification of the nature of the contract is finally a
convenient tool to decompose constraints, requirements or rules and is
probably business dependent.  In fact, the classification is just a
tool that helps improve the trust we can have in the coverage (not to
say the completeness) of a contract-based specification.

\section{Conclusion}

Beyond the primary idea of Design by Contracts~\cite{computer:1992}
which was a pragmatic and systematic way of using pre/post conditions
in a functional (or object-oriented) world, the idea to generalize the
use of contracts to components leads to many interesting works.

Contracts are a natural way to decompose and structure
specifications. It appears that a contract can be attached to almost
any software entity. One of the main challenges is probably to deal
with the many kinds of contracts and of contracts checkers.
But even if contacts are widely used, they are not yet a
first-class entity in UML or even standard profiles.

A fundamental question that arises is the role of contracts in the
specification of software entities. Can we consider the set of
contracts as ``the'' specification of a service or a component?
Another important question is the life cycle of contracts. Some
contracts can be directly used and statically checked while others
need to be refined and give birth to monitors, checkers and
negotiation artifacts. The refinement process of contracts depends on
the level of contract and is tightly coupled to the software entity
development process itself with the underlying question; what has to
be exposed in the contract, what has to be hidden or, in other words,
what is the level of grayness of the specification.\footnote{On this
  topic, the WCSI paper~\cite{ouederni} provides an interesting view.}

Beyond the specification of a single entity, the contract concept is now central
to many software architectures. Yet many complex questions need to be
answered: which contract metamodels are relevant for a given system,
how can designers conciliate different metamodels stemming from
various parts that are combined to build a system, how can one
integrate dynamic contract negotiation, etc.

So, ten years later, the concept of contract still appears attractive but
has probably not reached a level of maturity that would allow a
large use in industry. Rendezvous in ten more years \ldots

\end{document}